\documentclass[11pt]{article}
\usepackage{amssymb,amsmath,amsfonts}
\usepackage{graphicx}
\usepackage{graphics}
\usepackage{eepic,epsfig}

\begin{document}

\title{Wightman function and scalar Casimir densities for a wedge with two
cylindrical boundaries}
\author{A. A. Saharian$^{1,2}$\thanks{%
E-mail: saharian@ictp.it } \thinspace and A. S. Tarloyan$^{1,3}$ \\
\\
\textit{$^1$Department of Physics, Yerevan State University} \\
\textit{0025 Yerevan, Armenia }\\
\textit{$^2$The Abdus Salam International Centre for Theoretical Physics} \\
\textit{34014 Trieste, Italy }\\
\textit{$^3$ Yerevan Physics Institute, Yerevan, Armenia}\\
\textit{0036 Yerevan, Armenia }}
\maketitle

\begin{abstract}
Wightman function, the vacuum expectation values of the field square and the
energy-momentum tensor are investigated for a massive scalar field with
general curvature coupling parameter inside a wedge with two coaxial
cylindrical boundaries. It is assumed that the field obeys Dirichlet
boundary condition on bounding surfaces. The application of a variant of the
generalized Abel-Plana formula enables to extract from the expectation
values the contribution corresponding to the geometry of a wedge with a
single shell and to present the interference part in terms of exponentially
convergent integrals. The local properties of the vacuum are investigated in
various asymptotic regions of the parameters. The vacuum forces acting on
the boundaries are presented as the sum of self-action and interaction
terms. It is shown that the interaction forces between the separate parts of
the boundary are always attractive. The generalization to the case of a
scalar field with Neumann boundary condition is discussed.
\end{abstract}

\bigskip


\bigskip

\section{Introduction}

The nontrivial properties of the vacuum state are among the most important
predictions in quantum field theory. These properties are manifested in the
response of the vacuum to external influences such as external fields. A
simple model of the influence is realized by imposing prescribed boundary
conditions on the field operator. The distortion of the spectrum for the
zero-point fluctuations of a quantum field by these conditions results in
the shifts in the vacuum expectation values of physical observables, such as
the vacuum energy density and stresses, and induces vacuum forces acting on
constraining boundaries. This is the well known Casimir effect (see \cite%
{Most97,Plun86,Bord01,Milt02} and references therein). The Casimir effect is
common to all systems characterized by fluctuating quantities and has
important implications on all scales, from cosmological to subnuclear. In
addition to its fundamental interest this effect also plays an important
role in the fabrication and operation of nano- and micro-scale mechanical
systems and has become an increasingly popular topic in quantum field theory.

An interesting topic in the investigations of the Casimir effect has always
been the dependence of the physical characteristics of the vacuum on the
geometry of constraining boundaries. Analytic results can usually be found
only for highly symmetric geometries including planar, spherically and
cylindrically symmetric boundaries. Recently exact results for the Casimir
force in geometries of a sphere and a cylinder above a plate are obtained in
\cite{Bulg06,Emig06} (see also \cite{Most07}). Aside from their own
theoretical and experimental interest, the problems with this type of
boundaries are useful for testing the validity of various approximations
used to deal with more complicated geometries. In the present paper we
consider a less symmetric exactly solvable geometry of boundaries which is a
combination of a wedge with coaxial cylindrical shells. The Casimir effect
for wedge-shaped regions is well investigated in literature \cite%
{Most97,jphy,Deutsch,brevikI,brevikII,Nest02}. For a conformally coupled
scalar and electromagnetic fields the vacuum expectation value of the
energy-momentum tensor inside the wedge is azimuthal symmetric. In
particular, the vacuum energy-momentum tensor is finite everywhere apart
points on the edge. This property is a direct consequence of the conformal
invariance in the corresponding problems and does not take place for a
non-conformally coupled scalar field. For a scalar field with an arbitrary
curvature coupling parameter satisfying Dirichlet boundary condition on the
wedge sides the vacuum energy-momentum tensor is evaluated in \cite%
{Reza02,Saha05cyl}. In addition to the azimuthal dependence this tensor,
unlike to the case of conformally coupled fields, is also non-diagonal with
nonzero azimuthal-radial off-diagonal component.

The investigations of quantum effects for cylindrical boundaries have
received a great deal of attention. In addition to traditional problems of
quantum electrodynamics under the presence of material boundaries, the
Casimir effect for cylindrical geometries is also important in the flux tube
models of confinement \cite{Fish87,Barb90} and for determining the structure
of the vacuum state in interacting field theories \cite{Ambj83}. The
calculation of the vacuum energy for the electromagnetic field with boundary
conditions defined on a cylinder turned out to be technically a more
involved problem than the analogous one for a sphere. First the Casimir
energy of an infinite perfectly conducting cylindrical shell has been
calculated in Ref. \cite{Dera81} by introducing ultraviolet cutoff and later
the corresponding result was derived by using other methods \cite%
{Milt99,Gosd98,Lamb99}. The local characteristics of the corresponding
electromagnetic vacuum such as energy density and vacuum stresses are
considered in \cite{Sah1cyl} for the interior and exterior regions of a
conducting cylindrical shell, and in \cite{Sah2cyl} for the region between
two coaxial shells (see also \cite{Saha00rev}). The electromagnetic vacuum
forces acting on the boundaries in the geometry of two cylinders are also
considered in Refs. \cite{Mazz02}. In Ref. \cite{Rome01} scalar vacuum
densities and the zero-point energy for general Robin boundary condition on
a cylindrical surface in arbitrary number of spacetime dimensions are
studied for massive scalar field with general curvature coupling parameter.
The corresponding problem for the geometry of two coaxial cylindrical shells
is considered in \cite{Saha06cyl}. A large number of papers is devoted to
the investigation of the various aspects of the Casimir effect for a
dielectric cylinder (see, for instance, \cite{Milt02,Nest04} and references
therein).

In the geometry of a wedge with coaxial cylindrical boundary the modes are
still factorizable for both scalar and electromagnetic fields and the
corresponding problems are exactly solvable. The total Casimir energy of a
semi-circular infinite cylindrical shell with perfectly conducting walls is
considered in \cite{Nest01} by using the zeta function technique. For a
scalar field with an arbitrary curvature coupling parameter obeying
Dirichlet boundary condition the Wightman function, the vacuum expectation
values of the field square and the energy-momentum tensor in the geometry of
a wedge with an arbitrary opening angle and with a cylindrical boundary are
investigated in \cite{Reza02,Saha05cyl}. The corresponding Casimir densities
for the electromagnetic field with perfect conductor boundary conditions on
bounding surfaces are considered in \cite{Saha07El}. The closely related
problem with a cylindrical shell in the geometry of a cosmic string is
discussed in \cite{Beze06Sc,Beze07El} for scalar and electromagnetic fields.
In both scalar and electromagnetic cases the application of a variant of the
generalized Abel-Plana formula \cite{Saha00rev} enables to extract from the
vacuum expectation values the parts corresponding to the geometry of a wedge
without the cylindrical shell and to present the shell induced parts in
terms of rapidly converging integrals. This geometry is also interesting
from the point of view of general analysis for surface divergences in the
expectation values of local physical observables for boundaries with
discontinuities. The nonsmoothness of the boundary generates additional
contributions to the heat kernel coefficients (see, for instance, the
discussion in \cite{Nest04,Apps98,Dowk00,Nest03} and references therein).

In this paper we investigate one-loop vacuum quantum effects for a scalar
field in the geometry of a wedge with two coaxial cylindrical shells
assuming Dirichlet boundary condition on bounding surfaces. This geometry
generalizes various special cases previously considered in literature for
wedge-shaped and cylindrical boundaries. In addition, we also study the role
of nonzero mass of the field quanta. The presence of boundaries eliminates
the translational invariance and as a result the properties of the vacuum
are nonuniform. The most important quantities characterizing the local
properties of the vacuum are the expectation values of the field square and
the energy-momentum tensor. In addition to describing the physical structure
of the quantum field at a given point, the energy-momentum tensor acts as
the source of gravity in the Einstein equations. It therefore plays an
important role in modelling a self-consistent dynamics involving the
gravitational field. As the first step for the investigation of vacuum
densities we evaluate the positive frequency Wightman function. This
function gives comprehensive insight into vacuum fluctuations and determines
the response of a particle detector of the Unruh-DeWitt type. Having the
vacuum energy-momentum tensor we can derive the vacuum forces acting on
constraining boundaries evaluating the vacuum stresses at points on the
bounding surfaces. As we will see below, in the geometry under consideration
these forces are position dependent on the boundary and cannot be obtained
by the global method using the total Casimir energy (on the advantages of
the local method see also \cite{Acto96}). In the limiting case from the
results of the present paper the local vacuum densities are obtained for the
geometry of a rectangular waveguide (for the local analysis of quantum
fields confined in rectangular cavities see \cite{Acto96,Acto94,Acto95}).

The paper is organized as follows. The next section is devoted to the
evaluation of the Wightman function for a massive scalar field in the region
bounded by two cylindrical shells and by the wedge walls. This function is
decomposed into three parts: the first one corresponds to the geometry of a
wedge without cylindrical shells, the second one is induced by a single
cylindrical shell when the second shell is absent, and the third one is
induced by the presence of the second shell. By using the formula for the
Wightman function, in section \ref{sec:phi2EMT} the vacuum expectation
values of the field square and the energy-momentum tensor are evaluated and
their behavior is investigated in various asymptotic regions of the
parameters. In section \ref{sec:forces} we consider the vacuum forces acting
on bounding surfaces. For separate boundary elements these forces are
decomposed into self-action and interaction parts. The interaction forces
are investigated in detail and numerical examples are presented. On the
example of interaction forces we also demonstrate the limiting transition to
the geometry of a rectangular waveguide. Finally, the results are summarized
and discussed in section \ref{sec:Conclusion}.

\section{Wightman function}

\label{sec:WF}

\begin{figure}[tbph]
\begin{center}
\epsfig{figure=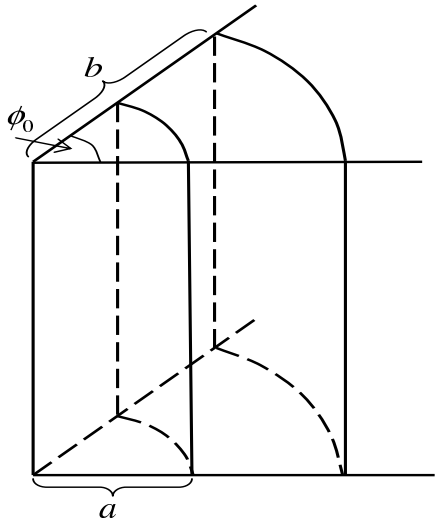,width=6.cm,height=7cm}
\end{center}
\caption{Geometry of a wedge with two coaxial cylindrical boundaries. }
\label{fig1}
\end{figure}

We consider a real scalar field $\varphi $ inside a wedge with opening angle
$\phi _{0}$ and with two coaxial cylindrical shells of radii $a$ and $b$, $%
a<b$ (see figure \ref{fig1}). For the field with curvature coupling
parameter $\xi $ the corresponding field equation has the form
\begin{equation}
\left( \nabla ^{i}\nabla _{i}+\xi R+m^{2}\right) \varphi \left( x\right) =0,
\label{fieldeq}
\end{equation}%
where $R$ is the curvature scalar for a $(D+1)$-dimensional background
spacetime, $\nabla _{i}$ is the covariant derivative operator. For special
cases of minimally and conformally coupled scalars one has $\xi =0$ and $\xi
=\xi _{D}\equiv (D-1)/4D$, respectively. Here we will assume that the
background spacetime is flat and, hence, in Eq. (\ref{fieldeq}) we have $R=0$%
. As a result the eigenmodes are independent of the curvature coupling
parameter. However, the local characteristics of the vacuum such as the
energy density and vacuum stresses depend on this parameter. In accordance
with the problem symmetry we will use cylindrical coordinates $\left( r,\phi
,z_{1},...,z_{N}\right) $, \ $N=D-2$, and will assume that the field obeys
Dirichlet boundary conditions on bounding surfaces:
\begin{equation}
\varphi |_{r=j}=\varphi |_{\phi =0}=\varphi |_{\phi =\phi _{0}}=0,\;j=a,b.
\label{boundcondD}
\end{equation}%
These boundary conditions modify the spectrum of the zero-point fluctuations
compared with the case of free space and change the physical properties of
the vacuum. Among the most important characteristics of the vacuum are the
expectation values of the field square and the energy-momentum tensor. These
expectation values can be obtained from two-point functions in the
coincidence limit. As a two-point function here we will consider the
positive frequency Wightman function $\left\langle 0|\varphi (x)\varphi
(x^{\prime })|0\right\rangle $, where $|0\rangle $ is the amplitude for the
vacuum state. This function also determines the response of Unruh-DeWitt
type particle detectors \cite{Birr82}. Here we consider the spatial region $%
0\leqslant \phi \leqslant \phi _{0}$, $a\leqslant r\leqslant b$. The
formulae for the regions $r\leqslant a$ and $r\geqslant b$ are obtained in
limiting cases. Note that by using the corresponding formulae we can discuss
various combinations of boundaries in the regions $0\leqslant \phi \leqslant
\phi _{0}$ and $\phi _{0}\leqslant \phi \leqslant 2\pi $. For example, we
can consider the situation with two cylindrical shells in the first region
and without shells in the second one.

By expanding the field operator and using the standard commutation
relations, the positive frequency Wightman function is presented as a sum
over the eigenmodes:%
\begin{equation}
\left\langle 0|\varphi (x)\varphi (x^{\prime })|0\right\rangle =\sum_{\alpha
}\varphi _{\alpha }(x)\varphi _{\alpha }^{\ast }(x),  \label{W1}
\end{equation}%
where $\left\{ \varphi _{\alpha }(x),\varphi _{\alpha }^{\ast }(x)\right\} $%
\ \ is a complete orthonormal set of positive and negative frequency
solutions to the field equation satisfying boundary conditions (\ref%
{boundcondD}). In the region between the cylindrical shells, $a\leqslant
r\leqslant b$, the eigenfunctions are specified by the set of quantum
numbers $\alpha =(n,\gamma ,\mathbf{k})$,$~\ n=1,2,\cdots $, and have the
form%
\begin{equation}
\varphi _{\alpha }(x)=\beta _{\alpha }g_{qn}(\gamma a,\gamma r)\sin (qn\phi
)\exp \left( i\mathbf{kr}_{\parallel }-i\omega t\right) ,  \label{eigfunc}
\end{equation}%
where%
\begin{equation*}
\omega =\sqrt{\gamma ^{2}+k_{m}^{2}},\;k_{m}^{2}=|\mathbf{k}|^{2}+m^{2},\
q=\pi /\phi _{0},
\end{equation*}%
and $\mathbf{r}_{\parallel }=\left( z_{1},...,z_{N}\right) $, $\mathbf{k}%
=(k_{1},\ldots ,k_{N})$, $-\infty <k_{j}<\infty $. In formula (\ref{eigfunc}%
) we have introduced the notation
\begin{equation}
g_{qn}(\gamma a,\gamma r)=Y_{qn}(\gamma a)J_{qn}(\gamma r)-J_{qn}(\gamma
a)Y_{qn}(\gamma r),  \label{gn}
\end{equation}%
with $J_{qn}(z)$ and $Y_{qn}(z)$ being the Bessel and Neumann functions. The
eigenfunctions $\varphi _{\alpha }(x)$ defined by (\ref{eigfunc}) satisfy
the boundary conditions on the inner shell and on the wedge sides. The
eigenvalues for the quantum number $\gamma $ are quantized by boundary
condition (\ref{boundcondD}) on the surface $r=b$ and are solutions of the
equation%
\begin{equation}
J_{qn}(\gamma a)Y_{qn}(\gamma b)-Y_{qn}(\gamma a)J_{qn}(\gamma b)=0.
\label{modeeq}
\end{equation}%
In the discussion below the corresponding positive roots we will denote by $%
\gamma a=\sigma _{qn,l}$, $l=1,2,\ldots $, assuming that they are arranged
in the ascending order, $\sigma _{qn,l}<\sigma _{qn,l+1}$.

The normalization coefficient $\beta _{\alpha }$ in (\ref{eigfunc}) is found
from the standard orthonormality condition for the eigenfunctions:
\begin{equation}
\int d^{N}\mathbf{r}_{\parallel }\int_{a}^{b}dr\,r\int_{0}^{\phi _{0}}d\phi
\,\varphi _{\alpha }(x)\varphi _{\alpha ^{\prime }}^{\ast }(x)=\frac{1}{%
2\omega }\delta _{nn^{\prime }}\delta _{ll^{\prime }}\delta (\mathbf{k-k}%
^{\prime }).  \label{normcond}
\end{equation}%
By making use of the standard integral for cylinder functions (see, for
instance, \cite{Prud86}), one finds%
\begin{equation}
\beta _{\alpha }^{2}=\frac{\pi ^{2}q\gamma T_{qn}^{ab}(\gamma a)}{(2\pi
)^{D-1}\omega a},  \label{betalf}
\end{equation}%
with the notation%
\begin{equation}
T_{\nu }^{ab}(z)=\frac{z}{J_{\nu }^{2}(z)/J_{\nu }^{2}(\eta z)-1},\;\eta
=b/a.  \label{Tnab}
\end{equation}%
The substitution of eigenfunctions (\ref{eigfunc}) into mode-sum formula (%
\ref{W1}) leads to the following expression for the positive frequency
Wightman function%
\begin{eqnarray}
\left\langle 0|\varphi (x)\varphi (x^{\prime })|0\right\rangle &=&\frac{\pi
^{2}q}{a}\int d^{N}\mathbf{k}\sum_{n=1}^{\infty }\sum_{l=1}^{\infty }\frac{%
zg_{qn}(z,zr/a)g_{qn}(z,zr^{\prime }/a)}{(2\pi )^{D-1}\sqrt{z+k_{m}^{2}a^{2}}%
}  \notag \\
&&\times \sin (qn\phi )\sin (qn\phi ^{\prime })\exp (i\mathbf{k}\Delta
\mathbf{r}_{\parallel }-i\omega \Delta t)T_{qn}^{ab}(z)\big|_{z=\sigma
_{n,l}},  \label{W2}
\end{eqnarray}%
where $\Delta \mathbf{r}_{\parallel }=\mathbf{r}_{\parallel }-\mathbf{r}%
_{\parallel }^{\prime }$ and $\Delta t=t-t^{\prime }$. As the expressions
for the eigenmodes $\sigma _{n,l}$ are not explicitly known, formula (\ref%
{W2}) for the Wightman function is not convenient. In addition, the separate
terms in the sum are highly oscillatory for large values of quantum numbers.
For the further evaluation of the summation over $l$ we apply formula \cite%
{Saha00rev}
\begin{eqnarray}
\frac{\pi ^{2}}{2}\sum_{l=1}^{\infty }h(\sigma _{qn,l})T_{qn}^{ab}(\sigma
_{n,l}) &=&\int_{0}^{\infty }\frac{h(x)dx}{J_{qn}^{2}(x)+Y_{qn}^{2}(x)}
\notag \\
&&-\frac{\pi }{4}\int_{0}^{\infty }dx\,\Omega _{a,qn}(x,\eta x)\left[
h(xe^{\pi i/2})+h(xe^{-\pi i/2})\right] ,  \label{Abel}
\end{eqnarray}%
which is a direct consequence of the generalized Abel-Plana formula (for
applications of the generalized Abel-Plana formula in investigations of the
vacuum densities in the Casimir effect see also \cite{Saha07rev}). In (\ref%
{Abel})%
\begin{equation}
\Omega _{a,qn}(x,y)=\frac{K_{qn}(y)/K_{qn}(x)}{%
K_{qn}(x)I_{qn}(y)-K_{qn}(y)I_{qn}(x)},  \label{Oma}
\end{equation}%
and $I_{qn}(x)$, $K_{qn}(x)$ are the modified Bessel functions.

As a function $h(x)$ in summation formula (\ref{Abel}) we choose%
\begin{equation}
h(x)=\frac{xg_{qn}(x,xr/a)g_{qn}(x,xr^{\prime }/a)}{\sqrt{%
x^{2}+k_{m}^{2}a^{2}}}\exp (-i\Delta t\sqrt{x^{2}/a^{2}+k_{m}^{2}}).
\label{hx}
\end{equation}%
The corresponding conditions for this formula to be valid are satisfied if $%
r+r^{\prime }+|\Delta t|<2b$. In particular, this is the case in the
coincidence limit $t=t^{\prime }$ for the region under consideration. As a
result, the Wightman function is presented in the form%
\begin{eqnarray}
\left\langle 0|\varphi (x)\varphi (x^{\prime })|0\right\rangle &=&\frac{q}{%
2^{D-2}\pi ^{D-1}}\sum_{n=1}^{\infty }\sin (qn\phi )\sin (qn\phi ^{\prime
})\int d^{N}\mathbf{k}\,e^{i\mathbf{k}\Delta \mathbf{r}_{\parallel }}  \notag
\\
&&\times \Bigg\{\int_{0}^{\infty }dx\frac{h(x)/a}{J_{qn}^{2}(x)+Y_{qn}^{2}(x)%
}-\frac{2}{\pi }\int_{k_{m}}^{\infty }dx\frac{x\Omega _{a,qn}(ax,bx)}{\sqrt{%
x^{2}-k_{m}^{2}}}  \notag \\
&&\times G_{qn}(ax,rx)G_{qn}(ax,r^{\prime }x)\cosh (\Delta t\sqrt{%
x^{2}-k_{m}^{2}})\Bigg\},  \label{W3}
\end{eqnarray}%
where $h(x)$ is defined by (\ref{hx}) and we have introduced the notation
\begin{equation}
G_{qn}(x,y)=K_{qn}(x)I_{qn}(y)-I_{qn}(x)K_{qn}(y).  \label{Gnj}
\end{equation}%
In the limit $b\rightarrow \infty $ the second term in figure braces on the
right of (\ref{W3}) vanishes, whereas the first term does not depend on $b$.
It follows from here that the part with the first term presents the Wightman
function for the geometry of a wedge with a single cylindrical shell of
radius $a$. The corresponding problem for a massless scalar field is
investigated in \cite{Saha05cyl}. For points $r<b$ the second term in figure
braces on the right of (\ref{W3}) is finite in the coincidence limit and,
hence, the renormalization procedure for the VEVs of the field square and
the energy-momentum tensor is reduced to the corresponding procedure for the
geometry with a single shell. In addition, in the coincidence limit of the
arguments the $x$-integral in (\ref{W3}) is exponentially convergent in the
upper limit.

In formula (\ref{W3}), the part corresponding to the geometry with a single
cylindrical shell with radius $a$ can be further transformed by using the
identity%
\begin{eqnarray}
\frac{g_{qn}(x,xr/a)g_{qn}(x,xr^{\prime }/a)}{J_{qn}^{2}(x)+Y_{qn}^{2}(x)}
&=&J_{qn}(xr/a)J_{qn}(xr^{\prime }/a)-\frac{1}{2}\sum\limits_{\sigma =1}^{2}%
\frac{J_{qn}(x)}{H_{qn}^{(\sigma )}(x)}  \notag \\
&&\times H_{qn}^{(\sigma )}(xr/a)H_{qn}^{(\sigma )}(xr^{\prime }/a),
\label{iden1}
\end{eqnarray}%
where $H_{qn}^{(\sigma )}(x)$, $\sigma =1,2$, are the Hankel functions. In
the corresponding integral over $x$ with the second term on the right of (%
\ref{iden1}) we rotate the integration contour by the angle $\pi /2$ for $%
\sigma =1$ and by the angle $-\pi /2$ for $\sigma =2$. Due to the well known
properties of the Hankel functions, under the condition $r+r^{\prime
}-|\Delta t|>2a$, the integrals over the arcs of the circle with large
radius vanish, whereas the integrals over $(0,iak_{m})$ and $(0,-iak_{m})$
cancel out. Introducing the Bessel modified functions one obtains%
\begin{eqnarray}
\int_{0}^{\infty }dz\frac{h(x)/a}{J_{qn}^{2}(x)+Y_{qn}^{2}(x)}
&=&\int_{0}^{\infty }dx\,x\frac{J_{qn}(xr)J_{qn}(xr^{\prime })}{\sqrt{%
x^{2}+k_{m}^{2}}}\exp (-i\Delta t\sqrt{x^{2}+k_{m}^{2}})  \notag \\
&&-\frac{2}{\pi }\int_{k_{m}}^{\infty }dx\frac{xI_{qn}(ax)}{K_{qn}(ax)}\frac{%
K_{qn}(xr)K_{qn}(xr^{\prime })}{\sqrt{x^{2}-k_{m}^{2}}}\cosh (\Delta t\sqrt{%
x^{2}-k_{m}^{2}}).  \label{int1}
\end{eqnarray}%
By taking into account this relation, the Wightman function is presented in
the form%
\begin{eqnarray}
\left\langle 0|\varphi (x)\varphi (x^{\prime })|0\right\rangle
&=&\left\langle \varphi (x)\varphi (x^{\prime })\right\rangle
_{0}+\left\langle \varphi (x)\varphi (x^{\prime })\right\rangle _{a}-\frac{q%
}{2^{D-3}\pi ^{D}}\sum_{n=1}^{\infty }\sin (qn\phi )\sin (qn\phi ^{\prime
})\,  \notag \\
&&\times \int d^{N}\mathbf{k}e^{i\mathbf{k}\Delta \mathbf{r}_{\parallel
}}\int_{k_{m}}^{\infty }dx\,x\frac{\Omega _{a,qn}(ax,bx)}{\sqrt{%
x^{2}-k_{m}^{2}}}G_{qn}(ax,rx)G_{qn}(ax,r^{\prime }x)  \notag \\
&&\times \cosh (\Delta t\sqrt{x^{2}-k_{m}^{2}}).  \label{W4}
\end{eqnarray}%
In this formula,%
\begin{eqnarray}
\left\langle \varphi (x)\varphi (x^{\prime })\right\rangle _{0} &=&\frac{q}{%
2^{D-2}\pi ^{D-1}}\sum_{n=1}^{\infty }\sin (qn\phi )\sin (qn\phi ^{\prime
})\int d^{N}\mathbf{k}\,e^{i\mathbf{k}\Delta \mathbf{r}_{\parallel }}  \notag
\\
&&\times \int_{0}^{\infty }dx\,x\frac{J_{qn}(xr)J_{qn}(xr^{\prime })}{\sqrt{%
x^{2}+k_{m}^{2}}}\exp (-i\Delta t\sqrt{x^{2}+k_{m}^{2}}),  \label{W0}
\end{eqnarray}%
is the Wightman function for the wedge without cylindrical boundaries, and
\begin{eqnarray}
\left\langle \varphi (x)\varphi (x^{\prime })\right\rangle _{a} &=&-\frac{q}{%
2^{D-3}\pi ^{D}}\sum_{n=1}^{\infty }\sin (qn\phi )\sin (qn\phi ^{\prime
})\int d^{N}\mathbf{k}\,e^{i\mathbf{k}\Delta \mathbf{r}_{\parallel }}  \notag
\\
&&\times \int_{k_{m}}^{\infty }dx\,x\frac{I_{qn}(ax)}{K_{qn}(ax)}\frac{%
K_{qn}(xr)K_{qn}(xr^{\prime })}{\sqrt{x^{2}-k_{m}^{2}}}\cosh (\Delta t\sqrt{%
x^{2}-k_{m}^{2}}),  \label{Wa}
\end{eqnarray}%
is the part of the Wightman function induced by a single cylindrical shell
with radius $a$ in the region $r>a$. Hence, the last term on the right of (%
\ref{W4}) is induced by the presence of the second shell with radius $b$.

An equivalent form for the Wightman function is obtained from (\ref{W4}) by
using the identity%
\begin{eqnarray}
&&\sum_{j=a,b}n_{j}\Omega _{j,qn}(ax,bx)G_{qn}(jx,xr)G_{qn}(jx,xr^{\prime })
\notag \\
&=&\frac{K_{qn}(bx)}{I_{qn}(bx)}I_{qn}(xr)I_{qn}(xr^{\prime })-\frac{%
I_{qn}(ax)}{K_{qn}(ax)}K_{qn}(xr)K_{qn}(xr^{\prime }),  \label{iden2}
\end{eqnarray}%
with the notations $n_{a}=1$, $n_{b}=-1$, and%
\begin{equation}
\Omega _{b,qn}(x,y)=\frac{I_{qn}(x)/I_{qn}(y)}{%
K_{qn}(x)I_{qn}(y)-K_{qn}(y)I_{qn}(x)}.  \label{Omb}
\end{equation}%
This leads to the following representation for the Wightman function%
\begin{eqnarray}
\left\langle 0|\varphi (x)\varphi (x^{\prime })|0\right\rangle
&=&\left\langle \varphi (x)\varphi (x^{\prime })\right\rangle
_{0}+\left\langle \varphi (x)\varphi (x^{\prime })\right\rangle _{b}-\frac{q%
}{2^{D-3}\pi ^{D}}\sum_{n=1}^{\infty }\sin (qn\phi )\sin (qn\phi ^{\prime
})\,  \notag \\
&&\times \int d^{N}\mathbf{k}e^{i\mathbf{k}\Delta \mathbf{r}_{\parallel
}}\int_{k_{m}}^{\infty }dx\,x\frac{\Omega _{b,qn}(ax,bx)}{\sqrt{%
x^{2}-k_{m}^{2}}}G_{qn}(bx,xr)G_{qn}(bx,xr^{\prime })  \notag \\
&&\times \cosh (\Delta t\sqrt{x^{2}-k_{m}^{2}}).  \label{W5}
\end{eqnarray}%
In this formula,%
\begin{eqnarray}
\left\langle \varphi (x)\varphi (x^{\prime })\right\rangle _{b} &=&-\frac{q}{%
2^{D-3}\pi ^{D}}\sum_{n=1}^{\infty }\sin (qn\phi )\sin (qn\phi ^{\prime
})\int d^{N}\mathbf{k}\,e^{i\mathbf{k}\Delta \mathbf{r}_{\parallel }}  \notag
\\
&&\times \int_{k_{m}}^{\infty }dx\,x\frac{K_{qn}(bx)}{I_{qn}(bx)}\frac{%
I_{qn}(xr)I_{qn}(xr^{\prime })}{\sqrt{x^{2}-k_{m}^{2}}}\cosh (\Delta t\sqrt{%
x^{2}-k_{m}^{2}})  \label{Wb}
\end{eqnarray}%
is the part induced by a single cylindrical shell of radius $b$ in the
region $r<b$ and the last term on the right is induced by the presence of
the second shell. Note that formulae (\ref{Wa}) and (\ref{Wb}) are related
by the interchange $a\rightleftarrows b$, $I_{n}\rightleftarrows K_{n}$. For
a massless scalar field these formulae are derived in \cite{Saha05cyl}.

\section{ VEVs of the field square and the energy-momentum tensor}

\label{sec:phi2EMT}

\subsection{Field square}

In this section we consider the VEVs for the field square and the
energy-momentum tensor in the region between the cylindrical shells. The VEV
of the field square is obtained from the Wightman function in the
coincidence limit of the arguments. In this limit and for points away from
the boundaries the divergences are contained in the term $\left\langle
\varphi (x)\varphi (x^{\prime })\right\rangle _{0}$ only. The corresponding
renormalization procedure is realized by subtracting the part for the
Minkowskian spacetime without boundaries. By using decompositions (\ref{W4})
and (\ref{W5}) for the Wightman function and taking the coincidence limit of
the arguments, for the renormalized VEV of the field square one finds%
\begin{equation}
\langle \varphi ^{2}\rangle _{\mathrm{ren}}=\langle \varphi ^{2}\rangle _{0,%
\mathrm{ren}}+\langle \varphi ^{2}\rangle _{j}+\langle \varphi ^{2}\rangle
_{jj^{\prime }},  \label{VEVphi2}
\end{equation}%
where $j^{\prime }=a$ ($b$) for $j=b$ ($a$) and the last term on the right
is given by the formula
\begin{eqnarray}
\langle \varphi ^{2}\rangle _{jj^{\prime }} &=&-2qA_{D}\sum_{n=1}^{\infty
}\sin ^{2}(qn\phi )  \notag \\
&&\times \int_{m}^{\infty }dx\,x\left( x^{2}-m^{2}\right) ^{\frac{D-3}{2}%
}\Omega _{j,qn}(ax,bx)G_{qn}^{2}(jx,rx),  \label{phi21}
\end{eqnarray}%
with the notation%
\begin{equation}
A_{D}=\frac{2^{2-D}}{\pi ^{(D+1)/2}\Gamma ((D-1)/2)}.  \label{AD}
\end{equation}%
To obtain this result we have used the formula%
\begin{equation}
\int_{0}^{\infty }dk\int_{k_{m}}^{\infty }dx\,\frac{k^{s}f(x)}{\sqrt{%
x^{2}-k_{m}^{2}}}=\frac{\pi ^{N/2}}{\Gamma \left( N/2\right) }B\left( \frac{%
N+s}{2},\frac{1}{2}\right) \int_{m}^{\infty }dx\,\left( x^{2}-m^{2}\right) ^{%
\frac{D-3}{2}}f(x),  \label{intform1}
\end{equation}%
where $B(x,y)$ is the Euler beta function. In formula (\ref{phi21}), the
term $\langle \varphi ^{2}\rangle _{0,\mathrm{ren}}$ is the renormalized VEV
for the geometry of a wedge without cylindrical shells and the term $\langle
\varphi ^{2}\rangle _{j}$ is induced by a single cylindrical shell of radius
$j$ when the second shell is absent. Hence, the last term is induced by the
second shell of radius $j^{\prime }$.

The formulae for single shell terms are directly obtained from (\ref{Wa})
and (\ref{Wb}) in the coincidence limit. By making use of formula (\ref%
{intform1}), in the case $j=a$ one finds
\begin{equation}
\langle \varphi ^{2}\rangle _{a}=-2qA_{D}\sum_{n=1}^{\infty }\sin
^{2}(qn\phi )\int_{m}^{\infty }dx\,x\left( x^{2}-m^{2}\right) ^{\frac{D-3}{2}%
}\frac{I_{qn}(ax)}{K_{qn}(ax)}K_{qn}^{2}(rx),  \label{phi2a}
\end{equation}%
and the formula for $\langle \varphi ^{2}\rangle _{b}$ is obtained from here
by the replacements $a\rightarrow b$, $I\rightleftarrows K$. Note that, as $%
\Omega _{j,qn}(x,y)>0$ for $x<y$, the both terms $\langle \varphi
^{2}\rangle _{j}$ and $\langle \varphi ^{2}\rangle _{jj^{\prime }}$ are
negative.\ For points away from the cylindrical shells the last two terms on
the right of formula (\ref{phi21}) are finite. Note that both single shell
and the second shell induced parts vanish on the wedge sides $\phi =0,\phi
_{0}$, $a<r<b$. The part $\langle \varphi ^{2}\rangle _{j}$ diverges on the
cylindrical surface $r=j$ with the leading term%
\begin{equation}
\langle \varphi ^{2}\rangle _{a}\approx -\frac{\Gamma \left( (D-1)/2\right)
}{(4\pi )^{(D+1)/2}|r-j|^{D-1}}.  \label{Phi2neara2}
\end{equation}%
for points with $|r/j-1|\ll |\sin \phi |,|\sin (\phi _{0}-\phi )|$. For
points near the edges $(r=j,\phi =0,\phi _{0})$ the leading terms in the
corresponding asymptotic expansions are the same as for the geometry of a
wedge with the opening angle $\phi _{0}=\pi /2$. The surface divergences in
the VEVs of local physical observables are well known in quantum field
theory with boundaries and are investigated for various types of bulk and
boundary geometries (see, for example, \cite%
{Deutsch,Birr82,Full89,Cand82,Kennedy,Baac86}).

The term $\langle \varphi ^{2}\rangle _{jj^{\prime }}$ in (\ref{VEVphi2})
vanishes on the shell $r=j$ and diverges on the shell $r=j^{\prime }$. The
corresponding surface divergences are the same as those for a single
cylindrical shell of radius $j^{\prime }$. It follows from here that if we
present the VEV of the field square in the form%
\begin{equation}
\langle \varphi ^{2}\rangle _{\mathrm{ren}}=\langle \varphi ^{2}\rangle _{0,%
\mathrm{ren}}+\sum_{j=a,b}\langle \varphi ^{2}\rangle _{j}+\Delta \langle
\varphi ^{2}\rangle ,  \label{interphi2}
\end{equation}%
then the interference term $\Delta \langle \varphi ^{2}\rangle $ is finite
everywhere. Let us consider the behavior of the interference part in
asymptotic regions of the parameters. In the limit $a\rightarrow 0$ for
fixed values $r$ and $b$, this term vanishes as $a^{2q}$. In the limit $%
b\rightarrow \infty $ and for a massless field the interference part tends
to zero like $1/b^{D+2q-1}$. In the same limit under the condition $mb\gg 1$
the interference part is suppressed by the factor $e^{-2mb}/b^{(D-1)/2}$.
For small values of the wedge opening angle one has $q\gg 1$ and, hence, the
order of the modified Bessel functions in the formulae for the VEVs is
large. By using the corresponding uniform asymptotic expansions (see, for
example, \cite{Abra64}) we can see that the main contribution comes from the
term with $n=1$ and from the lower limit of the $x$-integral. To the leading
order for the interference term we find%
\begin{equation}
\Delta \langle \varphi ^{2}\rangle \approx \frac{4q^{(D-1)/2}(a/b)^{2q}\sin
^{2}(q\phi )}{(2\pi )^{(D+1)/2}(b^{2}-a^{2})^{(D-1)/2}}.  \label{Deltaphi2q}
\end{equation}%
As we see, in this limit the interference part is exponentially suppressed.
For points not too close to the cylindrical shells, similar suppression
takes place for single shell induced parts.

\subsection{Vacuum energy-momentum tensor}

The VEV for the energy-momentum tensor is obtained by using the formula%
\begin{equation}
\langle 0|T_{ik}|0\rangle =\lim_{x^{\prime }\rightarrow x}\partial
_{i}\partial _{k}^{\prime }\langle 0|\varphi (x)\varphi (x^{\prime
})|0\rangle +\left[ \left( \xi -\frac{1}{4}\right) g_{ik}\nabla _{l}\nabla
^{l}-\xi \nabla _{i}\nabla _{k}\right] \langle 0|\varphi ^{2}|0\rangle .
\label{emtvev1}
\end{equation}%
Note that in this formula we have used the form of the metric
energy-momentum tensor which differs from the standard one by the term which
vanishes for the solutions of the field equation (see, for instance, \cite%
{Saha04EMT}). As in the case of the field square, for points away from the
boundaries the renormalization is realized by subtracting the part
corresponding to the Minkowski spacetime without boundaries. By using the
formulae for the Wightman function and for the VEV of the field square, for
the renormalized VEV we obtain%
\begin{equation}
\langle T_{i}^{k}\rangle _{\mathrm{ren}}=\langle T_{i}^{k}\rangle _{0,%
\mathrm{ren}}+\langle T_{i}^{k}\rangle _{j}+\langle T_{i}^{k}\rangle
_{jj^{\prime }},  \label{VEVemt}
\end{equation}%
where $j^{\prime }=a$ ($b$) for $j=b$ ($a$) and the non-zero components of
the last term on the right are given by the formulae (no summation over $i$)%
\begin{eqnarray}
\langle T_{i}^{i}\rangle _{jj^{\prime }} &=&\frac{1}{2}qA_{D}\sum_{n=1}^{%
\infty }\int_{m}^{\infty }dx\,x^{3}\left( x^{2}-m^{2}\right) ^{\frac{D-3}{2}%
}\Omega _{j,qn}(ax,bx)  \notag \\
&&\times \left\{
a_{i,qn}^{(+)}[G_{qn}(jx,rx)]-a_{i,qn}^{(-)}[G_{qn}(jx,rx)]\cos (2qn\phi
)\right\} ,  \label{vevemt_ii} \\
\langle T_{1}^{2}\rangle _{jj^{\prime }} &=&q^{2}A_{D}\sum_{n=1}^{\infty
}n\sin (2qn\phi )\int_{m}^{\infty }dx\,x^{2}(x^{2}-m^{2})^{\frac{D-3}{2}}
\notag \\
&&\times \Omega _{j,qn}(ax,bx)G_{qn}(jx,rx)\left[ \frac{2\xi }{rx}%
G_{qn}(jx,rx)+(1-4\xi )G_{qn}^{\prime }(jx,rx)\right] ,  \label{vevemt_12}
\end{eqnarray}%
with $G_{\nu }^{\prime }(x,y)=\partial _{y}G_{\nu }(x,y)$. In formula (\ref%
{vevemt_ii}) we have introduced notations%
\begin{eqnarray}
a_{i,l}^{(\pm )}[g(y)] &=&(4\xi -1)\left[ g^{\prime 2}(y)+\left( 1\pm \frac{%
l^{2}}{y^{2}}\right) g^{2}(y)\right] +2g^{2}(y)\frac{1-m^{2}r^{2}/y^{2}}{D-1}%
,  \label{ajpm} \\
a_{1,l}^{(\pm )}[g(y)] &=&g^{\prime 2}(y)+\frac{4\xi }{y}g(y)g^{\prime
}(y)-g^{2}(y)\left\{ 1\pm \left[ 1-4\xi (1\mp 1)\right] \frac{l^{2}}{y^{2}}%
\right\} ,  \label{ajpm1} \\
a_{2,l}^{(\pm )}[g(y)] &=&\left( 4\xi -1\right) \left[ g^{\prime
2}(y)+g^{2}(y)\right] -\frac{4\xi }{y}g(y)g^{\prime }(y)+\frac{l^{2}}{y^{2}}%
g^{2}(y)\left( 4\xi \pm 1\right) ,  \label{ajpm2}
\end{eqnarray}%
with $g(y)=G_{qn}(jx,y)$ and in (\ref{ajpm}) $i=0,3,\ldots ,D$. In
particular, for the vacuum energy density and stresses along directions
parallel to the cylinder axis we have the relations $\langle
T_{0}^{0}\rangle _{\mathrm{ren}}=\langle T_{3}^{3}\rangle _{\mathrm{ren}%
}=\ldots =\langle T_{D}^{D}\rangle _{\mathrm{ren}}$. This property is a
direct consequence of translation invariance of the problem along these
directions. In (\ref{VEVemt}) the term $\langle T_{i}^{k}\rangle _{j}$ is
induced by a single cylindrical surface with radius $j$ when the second
shell is absent and the term $\langle T_{i}^{k}\rangle _{jj^{\prime }}$ is
induced by the presence of the second shell. Note that the off-diagonal
component $\langle T_{1}^{2}\rangle _{jj^{\prime }}$ vanishes on the wedge
sides and on the cylindrical shell $r=j$. The formulae for the components $%
\langle T_{i}^{k}\rangle _{a}$ are obtained from (\ref{vevemt_ii}), (\ref%
{vevemt_12}) by the replacements%
\begin{equation}
\Omega _{j,qn}(ax,bx)\rightarrow
I_{qn}(ax)/K_{qn}(ax),\;G_{qn}(jx,rx)\rightarrow K_{qn}(rx).  \label{VEVemta}
\end{equation}%
The formulae for $\langle T_{i}^{k}\rangle _{b}$ are obtained from the
corresponding expressions for $\langle T_{i}^{k}\rangle _{a}$ by the
replacements $a\rightarrow b$, $I\rightleftarrows K$. Single shell parts in
both interior and exterior regions are investigated in \cite%
{Reza02,Saha05cyl} for a massless scalar field. These parts diverge on the
shell and for $|r/j-1|\ll |\sin \phi |,|\sin (\phi _{0}-\phi )|$ the leading
term in the corresponding asymptotic expansion is given by the formula (no
summation over $i$)
\begin{equation}
\langle T_{i}^{i}\rangle _{j}\approx \frac{D(\xi -\xi _{D})\Gamma \left(
(D+1)/2\right) }{2^{D}\pi ^{(D+1)/2}|r-j|^{D+1}},\quad i=0,2,\ldots ,D.
\label{T00asra2}
\end{equation}%
For the other components to the leading order one has $\langle
T_{1}^{1}\rangle _{j}\sim \langle T_{2}^{1}\rangle _{j}\sim |r-j|^{-D}$.

As in the case of the field square, the VEV\ of the energy-momentum tensor
can be presented in the form%
\begin{equation}
\langle 0|T_{i}^{k}|0\rangle =\langle T_{i}^{k}\rangle
_{0}+\sum_{j=a,b}\langle T_{i}^{k}\rangle _{j}+\Delta \langle
T_{i}^{k}\rangle ,  \label{Tikren}
\end{equation}%
where the surface divergences are contained in the single shell parts only
and the interference part is finite on the shells. The explicit formula for
the latter is obtained by subtracting from the last term on the right (\ref%
{vevemt_ii}) and (\ref{vevemt_12}) the corresponding single shell part. It
can be checked that the separate terms in formulae (\ref{interphi2}) , (\ref%
{Tikren}) satisfy the standard trace relation%
\begin{equation}
T_{i}^{i}=D(\xi -\xi _{D})\nabla _{i}\nabla ^{i}\varphi ^{2}+m^{2}\varphi
^{2},  \label{trrel}
\end{equation}%
and the continuity equation $\nabla _{i}T_{k}^{i}=0$. For the geometry under
consideration the latter takes the form%
\begin{eqnarray}
\partial _{r}\left( rT_{2}^{1}\right) +r\partial _{\phi }T_{2}^{2} &=&0,
\label{conteq1} \\
\partial _{r}\left( rT_{1}^{1}\right) +r\partial _{\phi }T_{1}^{2}
&=&T_{2}^{2}.  \label{conteq2}
\end{eqnarray}%
The behavior of the VEV for the energy-momentum tensor in the asymptotic
regions of the parameters is investigated in the way similar to that used
for the field square. In the limit $a\rightarrow 0$ the main contribution
comes from the term with $n=1$ and the interference part behaves as $a^{2q}$%
. For large values of the radius of the exterior shell, $b\rightarrow \infty
$, this part vanishes as $e^{-2mb}/b^{(D-1)/2}$ for a massive field and like
$1/b^{D+2q-1}$ for a massless one. For large values of the parameter $q$,
the interference term in the VEV of the energy-momentum tensor is suppressed
by the factor $(a/b)^{2q}$.

In the discussion above we have considered a model where the physical
interactions are replaced by the imposition of boundary conditions on the
field for all modes. Of course, this is an idealization as real physical
interactions cannot constrain all the modes of a fluctuating quantum field
\cite{Deutsch,Cand82,Grah02}. In general, the physical quantities in
problems with boundary conditions can be classified into two main groups
(see also \cite{Jaff06}). The first group includes quantities which do not
contain surface divergences. For these quantities the renormalization
procedure is the same as in quantum field theory without boundaries and they
can be evaluated by boundary condition calculations. The contribution of the
higher modes into the boundary induced effects in these quantities is
suppressed by the parameters already present in the idealized model.
Examples of such quantities are the vacuum densities away from boundaries
and the interaction forces between disjoint bodies. For the quantities from
the second group, such as the vacuum densities on the boundary and the total
vacuum energy, the contribution of the arbitrary higher modes is dominant
and they contain divergences which cannot be eliminated by the standard
renormalization procedure of quantum field theory without boundaries. Of
course, the model where the physical interaction is replaced by the
imposition of boundary conditions on the field for all modes is an
idealization. The appearance of divergences in the process of the evaluation
of physical quantities of the second type indicates that more realistic
physical model should be employed for their evaluation. In literature on the
Casimir effect different field-theoretical approaches have been discussed to
extract the finite parts from the diverging quantities. However, in the
physical interpretation of these results it should be taken into account
that these terms are only a part of the full expression of the physical
quantity and the terms which are divergent in the idealized model can be
physically essential and their evaluation needs a more realistic model. It
seems plausible that such effects as surface roughness, or the
microstructure of the boundary on small scales can introduce a physical
cutoff needed to produce finite values for surface quantities. Another
possibility, proposed in Refs. \cite{Grah02}, is to replace a boundary
condition by a renormalizable coupling between the fluctuating field and
non-dynamical smooth background field representing the material (for the
evaluation of the vacuum energy in smooth background fields see also \cite%
{Bord96}). In this model the standard renormalization procedure of quantum
field theory without boundaries provides the finite result for the
quantities which are divergent in the boundary condition limit. An
alternative mechanism for introducing a cutoff which removes singular
behavior on boundaries is to allow the position of the boundary to undergo
quantum fluctuations \cite{Ford98}. Such fluctuations smear out the
contribution of the high frequency modes without the need to introduce an
explicit high frequency cutoff.

The main subject of the present paper is the investigation of the VEVs for
the field square and the energy-momentum tensor at points away from the
boundaries and the vacuum interaction forces between separate parts of
boundaries. In the scheme where a cutoff function is used instead of
point-splitting, these quantities are cutoff independent and fall into the
first group. They do not contain surface divergences and are completely
determined within the framework of standard procedure of quantum field
theory without boundaries. We expect that similar results would be obtained
in the model where instead of externally imposed boundary condition the
fluctuating field is coupled to a smooth background potential that
implements the boundary condition in a certain limit \cite{Grah02}.

\section{Vacuum interaction forces}

\label{sec:forces}

In this section we investigate the vacuum forces acting on the bounding
surfaces due to the presence of the second cylindrical shell. First of all
let us consider the forces acting on the wedge sides. These forces are
determined by the $_{2}^{2}$-component of the energy-momentum tensor
evaluated for $\phi =0,\phi _{0}$. Note that the off-diagonal components $%
\langle T_{1}^{2}\rangle _{j}$ and $\langle T_{1}^{2}\rangle _{jj^{\prime }}$
vanish on the wedge sides and, hence do not contribute to the force. The
corresponding effective pressure is presented in the form%
\begin{equation}
p_{2}=p_{2,\mathrm{wedge}}+p_{2,\mathrm{cyl}},  \label{p2wcyl}
\end{equation}%
where $p_{2,\mathrm{wedge}}$ is the vacuum effective pressure on the wedge
side when the cylindrical shells are absent and the part $p_{2,\mathrm{cyl}}$
is induced by the shells. For a conformally coupled massless scalar in $D=3$
one has%
\begin{equation}
p_{2,\mathrm{wedge}}=-\frac{q^{4}-1}{480\pi ^{2}r^{4}}.  \label{p2wedge}
\end{equation}%
The corresponding force is attractive for $\phi _{0}<\pi $ and repulsive for
$\phi _{0}>\pi $. The second term on the right of (\ref{p2wcyl}) is
decomposed as

\begin{equation}
p_{2,\mathrm{cyl}}=p_{2,\mathrm{cyl}}^{(j)}+p_{2,\mathrm{cyl}}^{(jj^{\prime
})},  \label{p2}
\end{equation}%
where $p_{2,\mathrm{cyl}}^{(j)}=-\langle T_{2}^{2}\rangle _{j}|_{\phi =0}$
is the effective azimuthal pressure on the wedges induced by a single
cylindrical boundary with radius $j$, $j=a,b$, and $p_{2,\mathrm{cyl}%
}^{(jj^{\prime })}=-\langle T_{2}^{2}\rangle _{jj^{\prime }}|_{\phi =0}$ is
induced by the presence of the second cylindrical boundary. Substituting $%
i=2 $ and $\phi =0,\phi _{0}$ in the formulae for the VEVs of the
energy-momentum tensor from the previous section, for the forces induced by
the shells we find
\begin{eqnarray}
p_{2,\mathrm{cyl}}^{(a)} &=&-\frac{q^{3}A_{D}}{r^{2}}\sum_{n=1}^{\infty
}n^{2}\int_{m}^{\infty }dx\,x\left( x^{2}-m^{2}\right) ^{\frac{D-3}{2}}\frac{%
I_{qn}(ax)}{K_{qn}(ax)}K_{qn}^{2}(rx),  \label{p2cyla} \\
p_{2,\mathrm{cyl}}^{(jj^{\prime })} &=&-\frac{q^{3}A_{D}}{r^{2}}%
\sum_{n=1}^{\infty }n^{2}\int_{m}^{\infty }dx\,x\left( x^{2}-m^{2}\right) ^{%
\frac{D-3}{2}}\Omega _{j,qn}(ax,bx)G_{qn}^{2}(jx,rx).  \label{p2cyljjp}
\end{eqnarray}%
The expression for $p_{2,\mathrm{cyl}}^{(b)}$ is obtained from (\ref{p2cyla}%
) by the replacements $a\rightarrow b$, $I\rightleftarrows K$. Single shell
parts in the forces acting on the wedge sides, $p_{2,\mathrm{cyl}}^{(j)}$,
are finite for all values $r$ except the points on the edge $r=j$. The
second shell-induced part, $p_{2,\mathrm{cyl}}^{(jj^{\prime })}$, is finite
for all $r$ except the points on the edge $r=j^{\prime }$, $j^{\prime }=a,b$%
, $j^{\prime }\neq j$. Note that $p_{2,\mathrm{cyl}}^{(jj^{\prime })}=0$ for
$r=j$. The integrands in (\ref{p2cyla}) and (\ref{p2cyljjp}) are positive
and, hence, the corresponding vacuum forces are attractive. As before we can
write%
\begin{equation}
p_{2,\mathrm{cyl}}=\sum_{j=a,b}p_{2,\mathrm{cyl}}^{(j)}+\Delta p_{2,\mathrm{%
cyl}},  \label{p2interf}
\end{equation}%
where the interference part $\Delta p_{2,\mathrm{cyl}}$ is finite for all
values $a\leqslant r\leqslant b$. As it follows from (\ref{p2cyla}), (\ref%
{p2cyljjp}), the corresponding forces do not depend on the curvature
coupling parameter.

\begin{figure}[tbph]
\begin{center}
\epsfig{figure=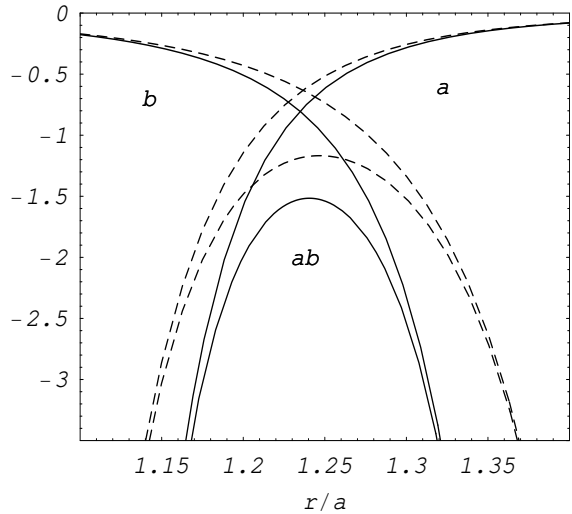,width=8.cm,height=7cm}
\end{center}
\caption{Vacuum forces acting on the wedge sides due to the
presence of cylindrical shells for $D=3$ massless scalar field:
$a^{4}p_{2,\mathrm{cyl}}^{(j)}$ and $a^{4}p_{2,\mathrm{cyl}}$. The
full (dashed) curves correspond to the wedge with $\protect\phi
_{0}=\protect\pi /2$  ($\protect\phi _{0}=3\protect\pi /2$) and we
have taken
$b/a=1.5$. The curves a (b) correspond to the effective pressure $a^{4}p_{2,%
\mathrm{cyl}}^{(a)}$ ($a^{4}p_{2,\mathrm{cyl}}^{(b)}$)\ when the shell with
radius $a$ ($b$) is present only, and the curves ab are for the effective
pressure $a^{4}p_{2,\mathrm{cyl}}$ when the both shells are present.}
\label{fig2}
\end{figure}

In the limit $a\rightarrow 0$ the main contribution into $p_{2,\mathrm{cyl}%
}^{(a)}$ and $\Delta p_{2,\mathrm{cyl}}$ comes from the term with $n=1$ and
these quantities behave like $a^{2q}$. In the limit $b\rightarrow \infty $
and for a massive scalar field the parts $p_{2,\mathrm{cyl}}^{(b)}$ and $%
\Delta p_{2,\mathrm{cyl}}$ are exponentially suppressed by the factor $%
e^{-2mb}$. In the same limit and for a massless field the main contribution
comes from the summand with $n=1$ and these parts behave as $1/b^{D+2q-1}$.
Now we consider the forces acting on the wedge sides in the limit of small
values of the opening angle when the parameter $q$ is large, $q\gg 1$. In
this limit the order of the modified Bessel functions is large and we can
use the uniform asymptotic expansions for these functions. By using these
expansions, it can be seen that the main contribution comes from the $n=1$
term and from the lower limit of the integral. To the leading order we find%
\begin{equation}
p_{2,\mathrm{cyl}}^{(j)}\approx -\frac{q^{(D+3)/2}\exp [-2q|\ln (j/r)|]}{%
(2\pi )^{(D+1)/2}r^{2}|r^{2}-j^{2}|^{(D-1)/2}}.  \label{p2jcylq}
\end{equation}%
In the similar way, for the interference part of the force one has:%
\begin{equation}
\Delta p_{2,\mathrm{cyl}}\approx \frac{2q^{(D+3)/2}(a/b)^{2q}}{(2\pi
)^{(D+1)/2}r^{2}(b^{2}-a^{2})^{(D-1)/2}}.  \label{Deltp2q}
\end{equation}%
In figure \ref{fig2} we have plotted the quantities $a^{4}p_{2,\mathrm{cyl}%
}^{(j)}$, $j=a,b$, and $a^{4}p_{2,\mathrm{cyl}}$ as functions of $r/a$ for $%
D=3$ massless scalar field. The graphs are given for the wedges with $\phi
_{0}=\pi /2$ (full curves) and $\phi _{0}=3\pi /2$ (dashed curves) and for $%
b/a=1.5$.

Now we turn to the interaction forces acting on the cylindrical boundaries.
These forces are determined by the $_{1}^{1}$-component of the
energy-momentum tensor evaluated on the corresponding surfaces. Similar to
the previous case, the effective pressure on the cylindrical shell $r=j$ is
presented as the sum
\begin{equation}
p^{(j)}=p_{1}^{(j)}+p^{(jj^{\prime })},  \label{p1j}
\end{equation}%
where $p_{1}^{(j)}=-(\langle T_{1}^{1}\rangle _{0}+\langle T_{1}^{1}\rangle
_{j})|_{r=j}$ is the radial vacuum stress on the cylinder with the radius $j$
when the second cylinder is absent and $p^{(jj^{\prime })}=-\langle
T_{1}^{1}\rangle _{jj^{\prime }}|_{r=j}$ is the additional stress on this
cylindrical surface when the second cylinder is present. Note that the
off-diagonal component $\langle T_{1}^{2}\rangle _{jj^{\prime }}$ vanishes
on the shell $r=j$ and does not contribute to the force. The part $%
p_{1}^{(j)}$ includes the self-action force on the cylindrical shell and
belongs to the second group of quantities in the classification given in the
previous section. Its evaluation requires more realistic model for the
interaction of the quantum field. Unlike to the self-action force, the
interaction force given by the second term on the right of (\ref{p1j}) is
finite for all nonzero distances between the shells and can be evaluated by
boundary condition calculations. From the last term on the right of (\ref%
{vevemt_ii}) taking $i=1$ and $r=j$ one finds:%
\begin{equation}
p^{(jj^{\prime })}=-\frac{qA_{D}}{j^{2}}\sum_{n=1}^{\infty }\sin ^{2}(qn\phi
)\int_{m}^{\infty }dx\,x\left( x^{2}-m^{2}\right) ^{\frac{D-3}{2}}\Omega
_{j,qn}(ax,bx).  \label{Deltap1j}
\end{equation}%
From this formula we see that $p^{(jj^{\prime })}<0$ and the corresponding
forces are always attractive. The expression for the interaction forces
between the cylindrical shells can also be written in the form%
\begin{equation}
p^{(jj^{\prime })}=\frac{qn_{j}A_{D}}{j}\frac{\partial }{\partial j}%
\sum_{n=1}^{\infty }\sin ^{2}(qn\phi )\int_{m}^{\infty }dx\,x\left(
x^{2}-m^{2}\right) ^{\frac{D-3}{2}}\ln \left[ 1-\frac{I_{qn}(ax)K_{qn}(bx)}{%
I_{qn}(bx)K_{qn}(ax)}\right] ,  \label{pjint1}
\end{equation}%
where, as before, $n_{a}=1$, $n_{b}=-1$. As for the forces acting on the
wedge sides, the interaction forces do not depend on the curvature coupling
parameter.

\begin{figure}[tbph]
\begin{center}
\epsfig{figure=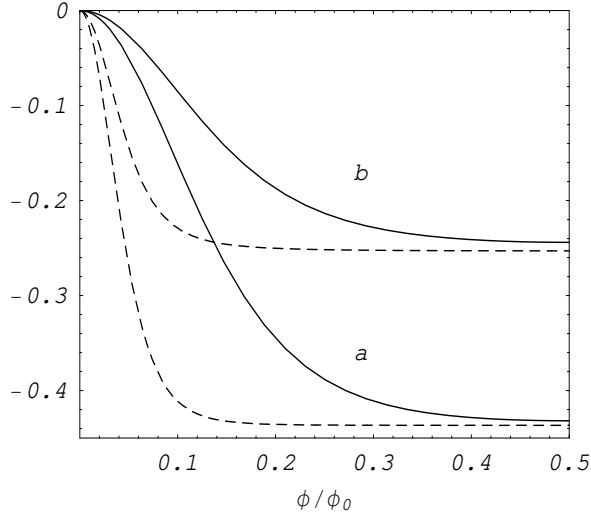,width=8.cm,height=7cm}
\end{center}
\caption{Vacuum forces acting on the cylindrical shell due to presence of
the second shell, $a^{4}p^{(jj^{\prime })}$ as functions on $\protect\phi /%
\protect\phi _{0}$, for $D=3$ massless scalar field. The full
(dashed) curves correspond to the wedge with $\protect\phi
_{0}=\protect\pi /2$ ($\protect\phi _{0}=3\protect\pi /2$) and in
both cases $b/a=1.5$. The curves a (b) correspond to the forces
acting on the shell with radius $a$ ($b$).} \label{fig3}
\end{figure}

Now we consider the behavior of the interaction forces in asymptotic regions
of the parameters. In the limit $a\rightarrow 0$ the main contribution in
the sum of formula (\ref{Deltap1j}) comes from the $n=1$ term and $%
j^{2}p^{(jj^{\prime })}\sim a^{2q}$. For large values of the exterior shell
radius, $b\rightarrow \infty $, and for a massive field the interaction
forces $p^{(jj^{\prime })}$ are suppressed by the factor $e^{-2mb}$. In the
same limit and for a massless field one has $j^{2}p^{(jj^{\prime })}\sim
1/b^{D+2q-1}$. For small values of the wedge opening angle, assuming that $%
q\gg 1$, in the way similar to that used for the estimation of the forces
acting on the wedge sides, one finds%
\begin{equation}
j^{2}p^{(jj^{\prime })}\approx -\frac{4q^{(D+3)/2}(a/b)^{2q}\sin ^{2}(q\phi )%
}{(2\pi )^{(D+1)/2}r^{2}(b^{2}-a^{2})^{(D-1)/2}}.  \label{pjjq}
\end{equation}%
In figure \ref{fig3} we have plotted the interaction forces acting on
cylindrical shells, $a^{4}p^{(jj^{\prime })}$, as functions of $\phi /\phi
_{0}$ for wedges with $\phi _{0}=\pi /2$ (full curves) and $\phi _{0}=3\pi
/2 $ (dashed curves) and for $b/a=1.5$ in the case of $D=3$ massless scalar
field. The curves a are for $p^{(ab)}$ and the curves b are for $p^{(ba)}$.

Note that in the geometry of two coaxial cylindrical shells without a wedge
the corresponding interaction forces are given by the formula \cite%
{Saha06cyl}
\begin{equation}
p^{(jj^{\prime })}=-\frac{A_{D}}{2j^{2}}\sideset{}{'}{\sum}_{n=0}^{\infty
}\int_{m}^{\infty }du\,u\left( u^{2}-m^{2}\right) ^{\frac{D-3}{2}}\Omega
_{j,n}(au,bu),  \label{pjj2cyl}
\end{equation}%
where the prime on the sum sign means that the term $n=0$ should be halved.
For $D=3$ massless scalar field and for $b/a=1.5$ from this formula we have $%
p^{(ab)}\approx -0.437/a^{4}$ and $p^{(ba)}\approx -0.254/a^{4}$. As it has
been shown in \cite{Saha06cyl}, the interaction forces (\ref{pjj2cyl}) can
also be obtained from the corresponding part in the total Casimir energy
differentiating over the radii of cylindrical shells. In the geometry under
consideration in the present paper the Casimir forces are position dependent
on the boundary and cannot be obtained by global methods using the total
Casimir energy.

In the limit $\phi _{0}\rightarrow 0$, $a,b\rightarrow \infty $, assuming
that $b-a\equiv L_{1}$ and $a\phi _{0}\equiv L_{2}$ are fixed, from the
formulae given above we obtain the corresponding results for the geometry of
a rectangular waveguide with sides $L_{1}$ and $L_{2}$. Here we discuss this
limiting transition for the case of the interaction forces $p^{(jj^{\prime
})}$. The consideration of the other quantities is done in the similar way.
In the limit under consideration the parameter $q$ is large and we can
replace the modified Bessel functions by the corresponding uniform
asymptotic expansions. By using these expansions it can be seen that to the
leading order we have%
\begin{equation}
\Omega _{j,\nu }(a\nu z,b\nu z)\approx \frac{2\nu \sqrt{1+a^{2}z^{2}}}{%
e^{2\nu \sqrt{1+a^{2}z^{2}}L_{1}/a}-1},\;\nu =qn.  \label{Omlim}
\end{equation}%
Introducing in (\ref{pjjq}) a new integration variable $z=x/qn$ and by
making use of (\ref{Omlim}), after some transformations, to the leading
order we find%
\begin{equation}
p^{(jj^{\prime })}\approx -\frac{2\pi A_{D}}{L_{1}^{D}L_{2}}%
\sum_{n=1}^{\infty }\sin ^{2}(\pi ny/L_{2})\int_{0}^{\infty }dx\,\frac{%
x^{D-2}\sqrt{x^{2}+c_{n}^{2}}}{e^{2\sqrt{x^{2}+c_{n}^{2}}}-1}%
,\;c_{n}^{2}=m^{2}L_{1}^{2}+(\pi nL_{1}/L_{2})^{2},  \label{pjjlim}
\end{equation}%
where $y=a\phi $. The expression on the right of this formula is the vacuum
interaction force per unit surface between the facets of the rectangular
parallelepiped separated by the distance $L_{1}$ and $y$ is the Cartesian
coordinate parallel to these facets. Other facets of the parallelepiped are
located at $y=0$ and $y=L_{2}$. Introducing in (\ref{pjjlim}) $y=y^{\prime
}+L_{2}/2$ and taking the limit $L_{2}\rightarrow \infty $ with fixed value $%
y^{\prime }$, from (\ref{pjjlim}) the vacuum forces for two infinite
parallel Dirichlet plates are obtained. Note that the local vacuum densities
for a quantum field confined within rectangular cavities are investigated in
\cite{Acto96,Acto94,Acto95} (for corresponding global quantities such as the
total Casimir energy see \cite{Most97,Milt02} and references therein).

\section{Conclusion}

\label{sec:Conclusion}

In this paper we have considered one-loop quantum vacuum effects for a
massive scalar field in the geometry of a wedge with two coaxial cylindrical
shells. We have assumed that the field satisfies Dirichlet boundary
condition on the bounding surfaces. This geometry generalizes various
special cases previously discussed in literature, including wedge-shaped
regions, cylindrical boundaries, and rectangular waveguides. The most
important local characteristics of the quantum vacuum are the VEVs for the
field square and the energy-momentum tensor. To evaluate these VEVs, as the
first step we construct the positive frequency Wightman function. The
corresponding eigensum contains a summation over the zeros of the
combination of Bessel and Neumann functions. The application of the
generalized Abel-Plana formula to the corresponding sum allows to present
the Wightman function in decomposed form given by formulae (\ref{W4}) and (%
\ref{W5}). In this representations the first term on the right is the
Wightman function for the wedge without cylindrical boundary, the term $%
\left\langle \varphi (x)\varphi (x^{\prime })\right\rangle _{j}$ is induced
by a single shell with radius $j$ when the second shell is absent, and the
last terms on the right are induced by the presence of the second shell. For
points away from the shells the last two terms are finite in the coincidence
limit and the renormalization is needed for the first term only. By taking
the coincidence limit, we have obtained similar representations for the VEVs
of the field square and the energy-momentum tensor, formulae (\ref{VEVphi2})
and (\ref{VEVemt}). More symmetric decompositions are given by formulae (\ref%
{interphi2}) and (\ref{Tikren}), where the last interference term is finite
everywhere including points on the shells. In the limit $a\rightarrow 0$ the
interference parts tends to zero like $a^{2q}$. For large values of the
exterior shell radius, $b\rightarrow \infty $, the interference terms in the
VEVs behave as $e^{-2mb}/b^{(D-1)/2}$ for a massive field and as $%
1/b^{D+2q-1}$ for a massless one. For a wedge with small opening angle, $%
q\gg 1$, the main contribution into the interference parts of the VEVs comes
from the summands with $n=1$ and these parts are suppressed by the factor $%
(a/b)^{2q}$.

In section \ref{sec:forces} we have considered the vacuum forces acting on
constraining boundaries. In the geometry under consideration these forces
are position dependent on the boundary and cannot be obtained by global
methods using the total Casimir energy. The forces acting on the wedge sides
are determined by the $_{2}^{2}$-component of the vacuum energy-momentum
tensor and are presented in the decomposed form (\ref{p2wcyl}). In this
representation the first term on the right determines the force when the
shells are absent and the second term is induced by the shells. In its turn
the latter is decomposed into a single shell and second shell induced parts
(see formula (\ref{p2})) given by formulae (\ref{p2cyla}), (\ref{p2cyljjp}).
Both these forces are always attractive and do not depend on the curvature
coupling parameter. Further we consider the forces acting on the cylindrical
shells. These force are presented in the form (\ref{p1j}) where the first
term on the right is the force acting on the cylindrical shell with radius $j
$ when the second shell is absent and the second term is induced by the
presence of the second shell. The latter, given by formula (\ref{Deltap1j}),
is always attractive and does not depend on the curvature coupling
parameter. For large values of the parameter $q$, this part is suppressed by
the factor $(a/b)^{2q}$. In the limit $\phi _{0}\rightarrow 0$, $%
a,b\rightarrow \infty $, assuming that $b-a$ and $a\phi _{0}$ are fixed,
from the results of the present paper we obtain the corresponding formulae
for the VEVs in the geometry of a rectangular waveguide. We have
demonstrated this on the example of the interaction force between the
cylindrical shells.

Note that we have considered quantities which are well defined
within the framework of standard renormalization procedure of
quantum field theory without boundaries. We expect that similar
results would be obtained from the model discussed in
\cite{Grah02} where instead of externally imposed boundary
condition the fluctuating field is coupled to a smooth background
potential that reproduces the boundary condition in a limiting
case. The generalization of the results in the present paper for a
scalar field with Neumann boundary conditions is straightforward.
For this case in the expressions (\ref{eigfunc}) of the
eigenfunctions the function $\cos (qn\phi )$ stands instead of
$\sin (qn\phi )$ and the quantum number $n$ takes the values
$0,1,2,\ldots $. The corresponding eigenvalues for $\gamma $ are
zeros of the function $J_{qn}^{\prime }(\gamma a)Y_{qn}^{\prime
}(\gamma b)-Y_{qn}^{\prime }(\gamma a)J_{qn}^{\prime }(\gamma b)$.
The formula for the summation over these zeros is given in
\cite{Saha00rev}. The formulae for the Wightman function and the
VEV of the field square in Neumann case are obtained from the
corresponding formulae for Dirichlet scalar by the
replacements $\sin (qn\phi )\rightarrow \cos (qn\phi )$, $%
I_{qn}(jx)\rightarrow I_{qn}^{\prime }(jx)$, $K_{qn}(jx)\rightarrow
K_{qn}^{\prime }(jx)$, $j=a,b$, and with the term $n=0$ included in the
summation. In the expressions for the VEVs of the energy-momentum tensor
this leads to the change of the sign for the second term in the figure
braces on the right of (\ref{vevemt_ii}) and to the change of the sign for
the off-diagonal component (\ref{vevemt_12}).

\section*{Acknowledgements}

AAS would like to acknowledge the hospitality of the Abdus Salam
International Centre for Theoretical Physics, Trieste, Italy. The work was
supported by the Armenian Ministry of Education and Science Grant No. 0124.

\end{document}